 \documentclass[preprint2]{aastex}

\newcommand{\mathsym}[1]{{}}
\newcommand{\unicode}[1]{{}}
 \usepackage{graphics}
 \usepackage{graphicx}
 \usepackage{epsfig}

\usepackage{amssymb}
\usepackage{amsmath}
\usepackage{aas_macros}

\newcommand{\lgbya}{\log\frac{b}{a}}

\shorttitle{Analysis of Linear Stability\dots Power-law Profile}
\shortauthors{Ram Kishor}

\begin{document}


\title{Periodic Orbits in the Generalized Photogravitational Chermnykh-Like Problem with Power-law Profile}


\author{Ram Kishor and Badam Singh Kushvah }

\affil{Department of   Applied  Mathematics,
Indian School of Mines, \\Dhanbad - 826004, Jharkhand, India}
%
\email{kishor.ram888@gmail.com; bskush@gmail.com} 

\begin{abstract}
 The orbits about Lagrangian equilibrium points are important  for scientific investigations. Since, a number of space missions have been completed and some are being proposed by various space agencies. In light of this, we consider a more realistic model in which a disk, with power-law density profile, is rotating around the common center of mass of the system. Then, we analyze the periodic motion in the neighborhood of Lagrangian equilibrium points for the value of mass parameter $0<\mu\leq \frac{1}{2}$. Periodic orbits of the infinitesimal mass in the vicinity of equilibrium are studied analytically and numerically. In spite of the periodic orbits, we have found some other kind of orbits like hyperbolic, asymptotic etc. The effect of radiation factor as well as oblateness coefficients  on the motion of infinitesimal mass in the neighborhood of equilibrium points are also examined. The stability criteria of the orbits examined by the help of Poincar\'{e} surfaces of section (PSS) and found that stability regions depend on the Jacobi constant as well as other parameters. \end{abstract}


\section{Introduction}
For the last few years, many authors have studied the periodic orbits of restricted three body problem (RTBP) due to its wide range of applications in space dynamics. One of them was \cite{Plummer1901MNRAS..62....6P} who studied the planar circular RTBP with arbitrary mass parameter $\mu$ and found that there are two families of periodic motion near the Lagrangian points. It was \cite{Szebehely1967torp.book.....S} who described complete results regarding the periodic motion of planar circular RTBP. A systematic classification of periodic orbits for $0<\mu\leq \frac{1}{2}$ in the neighborhood of Lagrangian points presented by \cite{Broucke1968QB401.B68......}. The periodic orbit about triangular point with mass parameter as critical mass of Routh studied by \cite{Meyerspringerlink:10.1007/BF01230325} whereas \cite{Markellosspringerlink:10.1007/BF01261880} investigated the problem numerically.  \cite{Ragos1991Ap&SS.182..313R} examined the periodic motion around the collinear equilibrium points in the 
photogravitational RTBP. However, \cite{Elipe1997CeMDA..68....1E} discussed the same motion by taking both primaries of RTBP as radiating. Further, \cite{Perdios2003Ap&SS.286..501P} studied critical symmetric periodic orbits of RTBP taking one primary as an oblate spheroid.  Whereas, \cite{Henonspringerlink:10.1023/A:1022518422926, Henonspringerlink:10.1007/s10569-005-3641-8} investigated new families of periodic orbits during the study of Hill's problem of the three body. On the other hand, \cite{Mittalspringerlink:10.1007/s10509-008-9942-0} examined the periodic motion of RTBP with oblateness by taking displacements along tangent and normal to the mobile coordinates. 

To analyze stability of the orbits, there are several methods like Lyapunov's, Poincar\'{e} Map or Poincar\'{e} surfaces of section \citep{Poincare1892QB351.P75......}, Henon's horizontal-vertical indices etc. 
 The PSS methods has been used by many authors like \citep{Ragos1997CeMDA..67..251R,1997WinterA&A...319..290W,2000WinterP&SS...48...23W},    to study the nature of   orbits around the equilibrium points in the restricted three body problem. \cite{SafiyaBeevi2011Ap&SS.333...37S} studied the periodic orbits in the Saturn-Titan problem using the numerical technique of PSS and found that the orbits around Saturn remain around it and their stability increases with the increase in the value of Jacobi constant. 
 
In the present paper, we have analyzed the periodic motion in the generalized photogravitational Chermnykh-like problem with power law density profile of disk which is rotating around the common center of mass of the system with mass ratio $0<\mu\leq \frac{1}{2}$. The Chermnykh-like problem has a number of applications in different areas such as celestial mechanics, chemistry, extra solar planetary system etc \citep{Gozdziewski1999CeMDA..75..251G, Strand1979JChPh..70.3812S, Rivera2000ApJ...530..454R, Jiang2001AA...367..943J}. This problem was first time studied by \cite{Chermnykh1987VeLen.......73C} latter many authors \citep{K.Gozdzieski1998CeMDA..70...41G, Papadakis2005Ap&SS.299..129P, Papadakis2005Ap&SS.299...67P, Jiang2006Ap&SS.305..341J, YehLCJiang2006, Kushvah2008Ap&SS.318...41K} have anaysed the different aspects of the problem theoretically as well as numerically. The proposed problem has been studied by \cite{Kushvah2012Ap&SS.337..115K} in the consequence of existence of equilibrium points and their linear stability. 

The numerical computation has been done with the help of  Mathematica$^{\textregistered}$  \cite{wolfram2003mathematica} software at necessary places by taking initial displacements $\xi_0=\eta_0=0.001$ and initial velocities $\dot{\xi_0}=\dot{\eta_0}=0.01$. The parametric values used in all numerical results are  $q_1=0.75$; $A_2=0.0025$; $c=1910.86$; $a=1$; $b=1.5$ and $h=10^{-4}$ unless otherwise stated. To find the PSS of the problem, we have used Event Locator method which is built into Mathematica$^{\textregistered}$ and works effectively as a controller. Since, the errors in a numerical method, generally, depend on step-size used which is controlled internally. Therefore, in a lower order integration routine error will be greater and hence more inaccuracy in results. As, a numerical method yields approximate results, consequently, there is an unavoidable error called as round off error which is the difference between approximate result and exact result. This error occurs due to finite precision representation of floating point numbers in the computer arithmetic. Since, we do not know the exact solution of the problem so, we can only control the errors by increasing working precision. On the other hand, a high working precision is unfavorable to computer and takes more time for execution of the results. Therefore, for a large time $t$ round off error will be large.
 
 The work is presented in six sections in which section(\ref{sec:model}) contains the mathematical formulation of the problem and equations of motion. Section(\ref{sec:eqvr}) includes the variational equations in the vicinity of equilibrium points. Section(\ref{sec:pocp}) and (\ref{sec:potp}) describe the periodic motions of the massless body in the neighborhood of collinear and triangular equilibrium points respectively. In section(\ref{sec:stb}), we have discussed the stability criteria with the help of PSS technique while section(\ref{sec:conc}) concludes the paper. 

\section{Mathematical Formulation of the Problem and Equations of Motion}
\label{sec:model}
It is supposed that $m_{1}$ and $m_{2}(m_1>m_2)$ be the masses of first (radiating body) and second ( oblate body) primaries respectively; $M_d$ be the total mass of the disk, rotating around the common center of mass of the system. The power law density profile of the disk having thickness $h\approx{10^{-4}}$ is $\rho(r)=\frac{c}{r^{p}}$, where $p$ is natural number(here we take $p=3$) and $c$ is a constant determined with the help of mass $M_d$. The forces, governing the motion of infinitesimal body $m$ are gravitational attractions due to both primaries as well as due to the disk. The radiation pressure force and oblateness coefficient are also taking into account. It is also assumed that the effect of the infinitesimal body on the motion of the remaining system is negligible. The unit of mass is taken in such a way that $G(m_1+m_2)={1}$; unit of distance is taken as the separation between the primaries whereas unit of time be the time period of the rotating frame $Oxyz$. The common center of mass of the 
primaries is taken as the origin $O$ of the rotating frame which is fixed relative to the inertial system. Let, $P(x,y,0)$, $A(-\mu,0,0)$ and $B(1-\mu,0,0)$ be the co-ordinates of infinitesimal body, first primary  and second primary respectively, with respect to the rotating frame, where $\mu=\frac{m_2}{m_1+m_2}$ is mass parameter.

 Therefore, the equations of motion of the infinitesimal body in $xy$-plane is written as  \citep{Kushvah2012Ap&SS.337..115K}:

\begin{eqnarray}
\ddot{x}-2n\dot{y}&=&\Omega_{x}, \label{eq:ux}\\
\ddot{y}+2n\dot{x}&=&\Omega_{y}, \label{eq:vx}  \end{eqnarray}
The potential function $\Omega$ is given as:\begin{eqnarray}
\Omega=\frac{n^{2}(x^{2}+y^{2})}{2}+\frac{(1-\mu)q_1}{r_1}+\frac{\mu}{r_2}+\frac{\mu{A_2}}{{2}r^{3}_2}-V, \label{eq:vv} \end{eqnarray}
where, $V$ is potential of the disk; $r=\sqrt{x^{2}+y^{2}}$; $r_1=\sqrt{(x+\mu)^{2}+y^{2}}$; $r_2=\sqrt{(x+\mu-1)^{2}+y^{2}}$; mass reduction factor $q_1=(1-\frac{F_p}{F_g})$, where $F_p$  and $F_g$ are the radiation pressure and gravitational attraction force of the radiating body respectively; $A_2=\frac{R^{2}_e-R^{2}_p}{5R^{2}}$ is the oblateness coefficient  \citep{McCuskey1963QB351.M3}, where $R_e$ and $R_p$ are the equatorial and polar radii of the same body respectively and $R$ is the distance between the primaries. The mean motion $n$ of the system  is given by $\sqrt{q_1+\frac{3}{2}A_2-2f_b(r)}$, where $f_b(r)$ is the gravitational force due to the disk. 

The  potential as well as gravitational force of the disk are given as \citep{Jiang2006Ap&SS.305..341J}: \begin{equation}
V=-4 \int_{r'}\frac{F({\zeta}) \rho(r')r'}{r+r'}dr',\label{eq:vbr}
\end{equation}
\begin{equation}
 f_b(r)=-2 \int_{r'}\frac{\rho(r')r'}{r}\left[\frac{E({\zeta})}{r-r'}+\frac{F({\zeta})}{r+r'}\right]dr',\label{eq:fbr}\end{equation}
where $F({\zeta})$ and $E({\zeta})$ are elliptic integrals of first and second kind respectively, $r'$ is the disk's reference radius and ${\zeta}=2\frac{\sqrt{rr'}}{r+r'}$. Now, with the help of expansion of the elliptic integrals for $a\leq r'\leq b$ in (\ref{eq:vbr}) and (\ref{eq:fbr}) and then selecting appropriate terms relative to $r$, the simplified form of $f_b(r)$ is given as:
\begin{eqnarray}
f_b(r)=-\pi c h\left[\frac{2(b-a)}{abr^2}+ \frac{3(\lgbya)}{8 r^{3}}\right], \end{eqnarray}
where, $a$ and $b$ are inner and outer radii of the disk respectively. It is assumed that the gravitational force $f_b(r)$ is radially symmetric  and hence  $\frac{x}{r}f_b(r)$ and $\frac{y}{r}f_b(r)$ are taken as the components of the force $f_b(r)$ along  $x$ and $y$ axes respectively.
From equation (\ref{eq:ux}) and (\ref{eq:vx}), we obtain the Jacobi  integral of the problem given as:
\begin{eqnarray}
&&C=-\dot{x}^{2}-\dot{y}^{2}+2U, \label{eq:ji} \end{eqnarray}
where  $C$ is Jacobi constant.

\section{Variational Equations in the Neighborhood of Equilibrium Points}
\label{sec:eqvr}
To analyze the possible motions of the restricted body in a small neighborhood of the equilibrium points $(x_e,y_e)$, we first make an infinitesimal change $\xi$ and $\eta$ in its coordinates i.e. $x=x_e+\xi$,\,$y=y_e+\eta$ such that the displacements 
\begin{eqnarray}
  \xi=P e^{\lambda{t}}, \  \eta=Q e^{\lambda{t}}, \label{eq:je}                                                                                                                                                                                                                                                          
 \end{eqnarray}                                                                                                                                                                                                                                                        are very small, where $P, Q$ and $\lambda$ are parameters to be determined. Substituting  these coordinates into equations (\ref{eq:ux}) and (\ref{eq:vx}), we get two differential equations of second order in variable $\xi$ and $\eta$ known as variational equations \citep{Murray2000ssd..book.....M}:
\begin{eqnarray}
\ddot{\xi}-2n\dot{\eta}={\xi}{\Omega^0_{xx}}+{\eta}{\Omega^0_{xy}},\nonumber\\
\ddot{\eta}+2n\dot{\xi}={\xi}{\Omega^0_{yx}}+{\eta}{\Omega^0_{yy}},\label{eq:var} \end{eqnarray} where superfix $0$ indicates the values are computed at the equilibrium point. Again, putting $\xi=P e^{\lambda{t}}$,\,$\eta=Q e^{\lambda{t}}$ into the equation (\ref{eq:var}) and simplifying them, we have
 \begin{eqnarray}
(\lambda^2-\Omega^0_{xx})P+(-2 n \lambda-\Omega^0_{xy})Q=0,\label{eq:pq1} \\
(2 n \lambda-\Omega^0_{yx})P+(\lambda^2-\Omega^0_{yy})Q=0. \label{eq:pq2}\end{eqnarray}
Now, the condition of nontrivial solution is that the determinant of the coefficients matrix of above system should be zero i.e.
\[
\begin{vmatrix}
\lambda^2-\Omega^0_{xx}& -2 n \lambda-\Omega^0_{xy} \\2 n \lambda-\Omega^0_{yx}& \lambda^2-\Omega^0_{yy}\\
\end{vmatrix}
=0.
\]
 So, from above we obtain a quadratic equation in $\lambda^{2}$ known as characteristic equation:
\begin{eqnarray}
&&\lambda^4+(4 n^2-\Omega^0_{xx}-\Omega^0_{yy}){\lambda^2}+\nonumber\\&&{(\Omega^0_{xx}}{\Omega^0_{yy}}-{\Omega^0}^2_{xy})=0.  \label{eq:ce} \end{eqnarray}
The four roots of characteristic equation (\ref{eq:ce}) play a crucial role to determine the form of orbits in the basin of equilibrium points.
Now, the second order derivatives of potential function $\Omega$ with respect to $x$ and $y$ are given as:
\begin{eqnarray}
\Omega_{xx}&=&n^2-\frac{q_1(1-\mu)}{r^3_1}\left[1-\frac{3(x+\mu)^2}{r^2_1}\right]\nonumber\\&&-\frac{\mu}{r^3_2}\left[1+\frac{3{A_2}}{2r^2_2}\right]-\frac{\pi c h}{r^3}\left[\frac{2(b-a)}{a b}\right.\nonumber\\&&\left.+\frac{7({\lgbya})}{r}\right]+\frac{3\mu(x+\mu-1)^2}{r^5_2}\left[1+\frac{5 A_2}{2r^2_2}\right]\nonumber\\&&+\frac{\pi c h  x^2}{r^5}\left[\frac{6(b-a)}{a b}-\frac{7({\lgbya})}{r}\right], \label{eq:oxx} 
\end{eqnarray}
\begin{eqnarray}
\Omega_{yy}&=&n^2-\frac{q_1(1-\mu)}{r^3_1}\left[1-\frac{3y^2}{r^2_1}\right]\nonumber\\&&-\frac{\mu}{r^3_2}\left[1+\frac{3{A_2}}{2r^2_2}\right]-\frac{\pi c h}{r^3}\left[\frac{2(b-a)}{a b}\right.\nonumber\\&&\left.+\frac{7({\lgbya})}{r}\right]+\frac{3y^2}{r^5_2}\left[1+\frac{5 A_2}{2r^2_2}\right]\nonumber\\&&+\frac{\pi c h  y^2}{r^5}\left[\frac{6(b-a)}{a b}-\frac{7({\lgbya})}{r}\right], \label{eq:oyy} \end{eqnarray}
\begin{eqnarray}
\Omega_{xy}&=&\frac{3q_1(1-\mu)(x+\mu) y}{r^5_1}\nonumber\\&&+\frac{3\mu(x+\mu-1) y}{r^5_2}\left[1+\frac{5 A_2}{2r^2_2}\right]\nonumber\\&& +\frac{\pi c h x y}{r^5}\left[\frac{6(b-a)}{a b}- \frac{7 (\lgbya)}{r}\right]. \label{eq:oxy} \end{eqnarray}

\section{Periodic Orbits in the Neighborhood of the Collinear Points}
\label{sec:pocp}
\begin{figure}[h] 
 \plotone{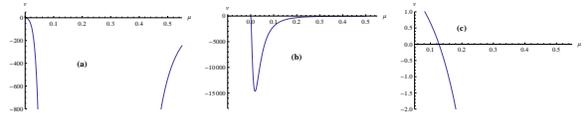}
 \caption{Values of $\nu=\Omega^0_{xx}\Omega^0_{yy}-(\Omega^0_{xy})^2$ with respect to mass parameter $\mu$: $(a) x_{L_1}=0.813609; \quad (b) x_{L_2}=1.05667$ and $(c) x_{L_3}=-0.823420$ \label{fig1:Cpp}}
 \end{figure}

 The values of second order derivatives of $\Omega$ at the collinear equilibrium points $(x_{L_i}, 0), i=1,2,3$ are given as:\begin{eqnarray}
\Omega^0_{xx}&=&n^2 +3B+\frac{3\mu {A_2}}{|(x_{L_i}+\mu-1)|^5}\nonumber\\&&-\frac{7\pi c h({\lgbya})}{4(x_{L_i})^4}\nonumber\\ 
\Omega^0_{yy}&=&n^2+B,\nonumber\\
 \Omega^0_{xy}&=&0,\end{eqnarray}                                                                                                                                                                                                                                                                                                                                                                                                                                                                                                                                                                                                                                                                                     where, \begin{eqnarray*}
 B&=&\frac{q_1(1-\mu)}{|(x_{L_i}+\mu)|^3}+\frac{\mu}{|(x_{L_i}+\mu-1)|^3}\nonumber\\&&+\frac{3\mu {A_2}}{2|(x_{L_i}+\mu-1)|^5}+ \frac{2\pi c h(b-a)}{a b|x_{L_i}|^3}\nonumber\\&&-\frac{7\pi c h({\lgbya})}{4(x_{L_i})^4}.                                                                                                                                                                                                                                                                                                                                                                 
                                                                                                                                                                                                                                                                                                                                                                  \end{eqnarray*}
Now, the last term of the characteristic equation (\ref{eq:ce}) are computed numerically by using the parametric values stated earlier and results are plotted in figure (\ref{fig1:Cpp}). It is found that 
\begin{eqnarray}
     \Omega^0_{xx}\Omega^0_{yy}-(\Omega^0_{xy})^2<0 \label{eq:enq}                                                                                                                                                                                                                                                           
                                                                                                                                                                                                                                                               \end{eqnarray}
   for all the collinear equilibrium points $x_{L_i}, i=1,2,3$ under the  
 condition of mass parameter $\mu$ whereas the above inequality holds for all values of $\mu$ \citep{Broucke1968QB401.B68......}. In other words in case of $L_1$ and $L_2$ above inequality  holds good for $0.01<\mu \leq \frac{1}{2}$ while in case of $L_3$ it is true for $0.13<\mu \leq \frac{1}{2}.$ 
Since, the zero degree term in the characteristic equation (\ref{eq:ce}) has always a negative sign  for these values of $\mu$. Hence, inequality (\ref{eq:enq}) guarantees that we always have two real roots of opposite sign in $\lambda^2$ under the restricted value of $\mu.$ That is \begin{eqnarray}
     \lambda^2= \alpha^2, \quad \lambda^2=-\beta^2.                                                                                                                                                                                                                                                  
                                                                                                                                                                                                                                                       \end{eqnarray}
Thus, the four roots $\lambda_j, j=1,2,3,4$ of the characteristic equation (\ref{eq:ce}) are given as\begin{eqnarray}
    &&\lambda_1= +\alpha, \quad \lambda_2=-\alpha, \nonumber\\&&  \lambda_3=+i\beta, \quad  \lambda_4=-i\beta. \label{eq:rot}                                                                                                     
   \end{eqnarray}
Hence, from equation (\ref{eq:je}) it is clear that the orbits corresponding to the roots $\lambda_1$ and $\lambda_2$ are of the exponential type. But orbits corresponding to roots $\lambda_3$ and $\lambda_4$ are periodic with period $\frac{2\pi}{\beta}.$

Now, the most general solution (which is of the complex type) of the variational equations (\ref{eq:var}) corresponding to the roots (\ref{eq:rot}) can be written as:
\begin{eqnarray}
\xi&=&P_1 e^{\alpha t}+P_2 e^{-\alpha t}+P_3 e^{i\beta t}+P_4 e^{-i\beta t}, \nonumber\\ 
\eta&=&Q_1 e^{\alpha t}+Q_2 e^{-\alpha t}+Q_3 e^{i\beta t}+\nonumber\\&&Q_4 e^{-i\beta t},\label{eq:xie}
\end{eqnarray}where $P_j, j=1,2,3,4$ are four arbitrary integration constants, whereas constants $Q_j, j=1,2,3,4$ are related to previous constants by the linear equation (\ref{eq:pq1}). That is
\begin{eqnarray}
 Q_j=\left(\frac{\lambda^{2}_j-\Omega^0_{xx}}{2\lambda_j}\right)P_j, j=1,2,3,4.
\end{eqnarray}
Let us suppose that 
\begin{eqnarray}
 &&Q_1=\gamma P_1, \quad Q_2=-\gamma P_2, \nonumber\\&& Q_3=i\delta P_3, \quad Q_4=-i\delta P_4, 
\end{eqnarray}where 
\begin{eqnarray}
\gamma&=&\frac{1}{2 \alpha}\left[\alpha^2-n^2 -3B-\frac{3\mu {A_2}}{|(x_{L_i}+\mu-1)|^5}\right.\nonumber\\&&\left.+\frac{7 \pi c h ({\lgbya})}{4(x_{L_i})^4}\right] \\
\delta&=&\frac{1}{2\beta}\left[\beta^2+n^2 +3B+\frac{3\mu {A_2}}{|(x_{L_i}+\mu-1)|^5}\right.\nonumber\\&&\left.-\frac{7 \pi c h ({\lgbya})}{4(x_{L_i})^4}\right]\label{eq:del} .
\end{eqnarray}
Then, the general solution (\ref{eq:xie}) takes the form
\begin{eqnarray}
\xi&=&P_1 e^{\alpha t}+P_2 e^{-\alpha t}+P_3 e^{i\beta t}+P_4 e^{-i\beta t}, \nonumber\\ 
\eta&=&\gamma \left( P_1 e^{\alpha t}-P_2 e^{-\alpha t}\right)\nonumber\\&&+i\delta \left(P_3e^{i\beta t}- P_4 e^{-i\beta t}\right),\label{eq:xien}
\end{eqnarray}
Also, the most general real solution of variational equations (\ref{eq:var}) can be written as:
 \begin{eqnarray}
 \xi&=&P_1 \cosh{\alpha t}+P_2 \sinh{\alpha t}+P_3 \cos{\beta t}+\nonumber\\&&P_4 \sin{\beta t}, \nonumber\\ 
\eta&=&\gamma \left(P_1 \sinh{\alpha t}+P_2 \cosh{\alpha t}\right)-\delta \left(P_3 \sin{\beta t}\right.\nonumber\\&&\left.+P_4 \cos{\beta t}\right),\label{eq:rxie}
 \end{eqnarray}where $P_j, j=1,2,3,4$ are arbitrary real integration constants. 
Now, taking appropriate initial conditions for the perturbed motion in the basin of collinear point, we have different type of orbits.

\subsection{Periodic Orbits}
\label{subsec:po}
It is supposed that initial displacements and velocities are $\xi_0$, $\eta_0$, $\Dot \xi_0$ and $\Dot \eta_0$ respectively. Then at $t=0$, we have from the equations (\ref{eq:xie}):
\begin{eqnarray}
&&\xi_0=P_1+P_2+P_3+P_4,\nonumber\\ 
&&\eta_0=\gamma \left(P_1-P_2\right) +i\delta \left(P_3-P_4\right),\nonumber\\
&&\Dot \xi_0=\alpha\left(P_1-P_2\right)+i\beta \left(P_3-P_4\right),\nonumber \\ 
&&\Dot \eta_0=\alpha \gamma \left(P_1+P_2\right)+i\beta \delta \left(P_3+P_4\right).\label{eq:cns}\end{eqnarray}
Again, from the equation (\ref{eq:xien}) it is clear that for periodic motion, $P_1$ and $P_2$ must be zero i.e. $P_1=P_2=0$ . Thus, from first two equations of (\ref{eq:cns}), we get \begin{eqnarray}
 P_3=\frac{\delta \xi_0-i\eta_0}{2 \delta}, \ P_4=\frac{\delta \xi_0+i\eta_0}{2 \delta}.\label{eq:p34}                                                                                                                                                   
                                                                                                                                                   \end{eqnarray}Using above, the equation (\ref{eq:xien}) reduces to following form:
\begin{eqnarray}
 \xi&=&\left(\frac{\delta \xi_0-i\eta_0}{2 \delta}\right) e^{i\beta t}+\left(\frac{\delta \xi_0+i\eta_0}{2 \delta}\right) e^{-i\beta t}, \nonumber\\ 
\eta&=&i\delta \left[\left(\frac{\delta \xi_0-i\eta_0}{2 \delta}\right) e^{i\beta t}-\right.\nonumber\\&&\left.\left(\frac{\delta \xi_0+i\eta_0}{2 \delta}\right) e^{-i\beta t}\right]\label{eq:xsc}
\end{eqnarray}
Now, simplifying these equation with the help of Euler's relations, we obtain
\begin{eqnarray}
 &&\xi=\xi_0 \cos{\beta t}+\frac{\eta_0}{\delta} \sin{\beta t}, \nonumber\\ 
&&\eta=\eta_0 \cos{\beta t}-\delta \xi_0 \sin{\beta t}.\label{eq:sct}
\end{eqnarray}
The parametric equations (\ref{eq:sct}) represent the periodic orbits of the infinitesimal mass in the neighborhood of collinear points $L_i$ with periods $T_i=\frac{2\pi}{\beta_{L_i}}, \ i=1,2,3$ which are depicted in figure (\ref{fig3:col}). 
 Now, eliminating parameter $t$ from these two equations, we get
\begin{eqnarray}
 \frac{\xi^2}{(\xi^2_0+\frac{\eta^2_0}{\delta^2})}+\frac{\eta^2}{(\delta^2 \xi^2_0+\eta^2_0)}=1. \label{eq:elp}\end{eqnarray}
which is the equation of an ellipse with center at collinear points $L_1$, $L_2$ or $L_3$ and axes parallel to $x$ and $y$ axis of the rotating frame. Also, the ellipse is bounded by rectangular region $\xi=\pm \sqrt{(\xi^2_0+\frac{\eta^2_0}{\delta^2})}$ and $\eta=\pm \sqrt{(\delta^2 \xi^2_0+\eta^2_0)}$, where $\delta$ defines the shape of the ellipse. The eccentricity of this ellipse (\ref{eq:elp}) is given as: 
\begin{eqnarray}
 \begin{cases}
e=(1-\frac{1}{\delta^2})^\frac{1}{2}; \ \delta^2>1,\\
e=(1-\delta^2)^\frac{1}{2}; \ 0<\delta^2<1.
 \end{cases}
\end{eqnarray}
\begin{figure}[h]
 \plotone{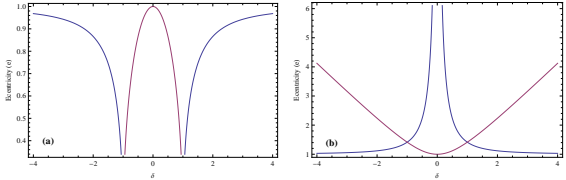}
 \caption{Variation of eccentricity \label{fig2:ecn}}
 \end{figure}
In figure (\ref{fig2:ecn}), frame $(a)$ shows that the eccentricity of ellipse lie in between interval $\left(0,\ 1 \right)$ and vary with the value of $\delta$. In other words, ellipticity of the orbit get increases by increasing the value of $\delta$ with condition $\delta^2>1$ whereas it decreases when $\delta^2<1$. Accordingly, the axes of the orbits get change with time. From equation (\ref{eq:del}), it clear that $\delta$ is the function of parameters $q_1, \ A_2$ and disk's radii and hence, the shape of orbit depends significantly on these parameters. Frame $(b)$, shows that variation of eccentricity of the hyperbolic orbit (\ref{eq:hyp}) with the value of $\delta.$

Alternatively, if we take $P_1=P_2=0$ in the real solution (\ref{eq:rxie}), we have
\begin{eqnarray}
 &&\xi=P_3 \cos{\beta t}+P_4 \sin{\beta t}, \nonumber\\ 
&&\eta=-\delta \left(P_3 \sin{\beta t}+P_4 \cos{\beta t}\right),\label{eq:rsc}
\end{eqnarray}which also represents a periodic motion.
Simplifying equation (\ref{eq:rsc}) by taking displacements  $\xi=\xi_0$, $\eta=\eta_0$ at $t=0$, we get
\begin{eqnarray}
 \xi=\xi_0 \cos{\beta t}; \quad \eta=-\delta \xi_0 \sin{\beta t};\label{eq:isc}
\end{eqnarray}which is a one parameter family of an ellipse having center at collinear points $L_1$, $L_2$ or $L_3$ and intersecting $\xi$-axis perpendicularly. Due to the positive value of constant $\delta$ motion on these ellipse will be retrograde.

In figure (\ref{fig3:col}), the frames $(a), \ (b)$ and $(c)$ indicate that periodic orbits in the neighborhood of collinear points $L_1, \ L_2$ and $L_3$ with periods $1.25876, \ 0.843075$ and $3.76437$ which are drawn at mass parameter $\mu=0.05, \ 0.05$ and $0.15$ respectively. It is seen that orbits get a deflection from their original path when we change the values of parameter $q_1$ or $A_2$. In other words, the effects of radiation pressure and oblateness  on the motion of infinitesimal mass are considerable. It is found that, generally the nature of the motion is unaffected but there is a miner change in time period of the orbits shown in Table \ref{tab:frst}. In this table, last three columns indicate the time periods of the periodic orbits in the vicinity of collinear points $L_1, \ L_2$ and $L_3$ respectively. It is clear that the time periods of the orbit decrease with increasing the value of $q_1$ or $A_2$. In other words, time periods of the orbits in the basin of collinear points decrease with 
an increase in radiation pressure. The oblateness coefficient also reduces  the time periods of orbits. That is, if $A_2$ increases then the  motion around equilibrium points becomes fast.
\begin{figure}[h]
 \plotone{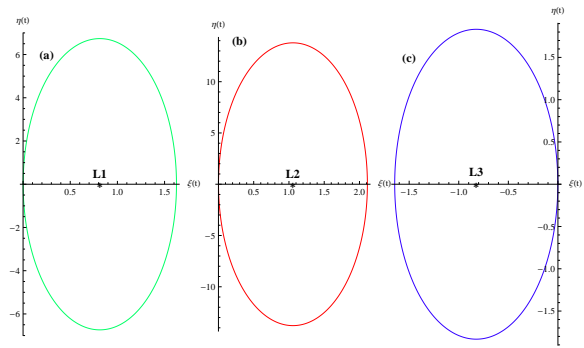}
 \caption{Periodic orbits in the neighborhood of collinear equilibrium points: $(a) x_{L_1}=0.813609; \quad (b) x_{L_2}=1.05667$ and $(c) x_{L_3}=-0.823420$ \label{fig3:col}}
 \end{figure}

\begin{table}
 \scriptsize
\caption{Effect of radiation factor and oblateness on the motion of infinitesimal mass\label{tab:frst}}
 \begin{center}
\begin{tabular}{|rrrr|}
\hline
 \bf{}&\bf{$T_{L_1}$}&\bf{$T_{L_2}$}&\bf{$T_{L_3}$}\\\hline
$q_1$&&$A_2=0.0025$&\\\hline
0.75&1.25876&0.843075&3.76437\\
0.85&1.25553&0.842186&3.53663\\
0.95&1.25230&0.841297&3.34299\\\hline
$A_2$&&$q_1=0.75$&\\\hline
0.0020&1.25878&0.843082&3.76995\\
0.0025&1.25876&0.843075&3.76437\\
0.0035&1.25871&0.843061&3.76065\\
0.0040&1.25869&0.843055&3.75881\\\hline
\end{tabular}
\end{center}
\end{table}

\subsection{Hyperbolic Orbits}
 \label{subsec:hpo}

Now, to discuss the unbounded motion of infinitesimal mass, we assume that the arbitrary constants $P_3$ and $P_4$ in equation (\ref{eq:xien}) are zero. Again, taking the same initial displacements and velocities as in case of periodic orbits. Then, from first two equations of (\ref{eq:cns}), we find the following 
\begin{eqnarray}
  P_1=\frac{\gamma \xi_0+\eta_0}{2 \gamma}, \quad P_2=\frac{\gamma \xi_0-\eta_0}{2 \gamma}.\label{eq:p12}
\end{eqnarray}
Substituting these values of arbitrary constants in the equation (\ref{eq:xie}), we get
\begin{eqnarray}
 \xi&=&\left(\frac{\gamma \xi_0+i\eta_0}{2 \gamma}\right) e^{\alpha t}+\left(\frac{\gamma \xi_0-i\eta_0}{2 \gamma}\right) e^{-\alpha t}, \nonumber\\ 
\eta&=&\gamma \left[\left(\frac{\gamma \xi_0+i\eta_0}{2 \gamma}\right) e^{\alpha t}-\right.\nonumber\\&&\left. \left(\frac{\gamma \xi_0-i\eta_0}{2 \alpha}\right) e^{i\alpha t}\right]\label{eq:xtc}
\end{eqnarray}
 Euler's relations reduces the equation (\ref{eq:xtc}) in following form
\begin{eqnarray}
 &&\xi=\xi_0 \cosh{\alpha t}+\frac{\eta_0}{\gamma} \sinh{\alpha t}, \nonumber\\ 
&&\eta=\gamma \xi_0 \sinh{\alpha t}+\eta_0 \cosh{\alpha t}.\label{eq:sch}
\end{eqnarray}The parametric equations (\ref{eq:sch}) represent path of the infinitesimal body in the vicinity of collinear points for the appropriate values of $\mu$ as described earlier. Eliminating parameter $t$, we have  
\begin{eqnarray}
 \frac{\xi^2}{(\xi^2_0-\frac{\eta^2_0}{\gamma^2})}-\frac{\eta^2}{(\gamma^2 \xi^2_0-\eta^2_0)}=1, \label{eq:hyp}\end{eqnarray}which is the equation of hyperbola with center at $L_1$, $L_2$ or $L_3$ and axes parallel to the coordinates axes of the rotating frame. The hyperbola (\ref{eq:hyp}) is bounded by the rectangular region $\xi=\pm \sqrt{(\xi^2_0-\frac{\eta^2_0}{\gamma^2})}$ and $\eta=\pm \sqrt{(\gamma^2 \xi^2_0+\eta^2_0)}$, where $\gamma$ defines the shape of the hyperbola. The eccentricity of this hyperbola is given as:
\begin{eqnarray}
 \begin{cases}
e=(1+\frac{1}{\gamma^2})^\frac{1}{2}; \ \gamma^2>1,\\
e=(1+\gamma^2)^\frac{1}{2}; \ 0<\gamma^2<1.
 \end{cases}
\end{eqnarray}

Alternatively, the hyperbolic motion in the vicinity of a collinear point can be examined by choosing the values of real arbitrary constants $P_3$ and $P_4$ in the equation (\ref{eq:rxie}) as equal to zero. If we changing the time origin then motion can be represented as:
\begin{eqnarray}
\xi=\xi_0 \cosh{\alpha t}; \quad \eta=\gamma \xi_0 \sinh{\alpha t};\label{eq:hsc} 
\end{eqnarray}which represent the one parameter family of hyperbolas with center at $L_1$, $L_2$ or $L_3$ and crossing the $\xi$-axis at right angle.

 \subsection{Asymptotic Orbits}
\label{subsec:asym}

The asymptotic motion is possible for the particular values of the real arbitrary constants in the equation (\ref{eq:rxie}). In other words, if we choose $P_3=P_4=0$ and $P_1=\pm P_2$ then the solution is written as:
\begin{eqnarray}
 \xi=\pm \xi_0 e^{\pm \alpha t}; \quad \eta=\pm \gamma \xi_0 e^{\pm \alpha t};\label{eq:as}
\end{eqnarray}which represent a rectilinear path of the infinitesimal body in the basin of the collinear points. This path may be considered as a hyperbola degenerated in its asymptotes. Plus minus sign in the equation (\ref{eq:as}), indicate that the infinitesimal body going away from or coming towards one of the collinear points $L_i, i=1,2,3$ asymptotically with time $t.$ For each collinear equilibrium point, there are four possible asymptotic orbits according to the signs of equation (\ref{eq:as}). Again, due to symmetry of the RTBP relative to $x$-axis, only two asymptotic orbits are really fundamental while other two are images of them. In other words
\begin{eqnarray}
 \xi=\xi_0 e^{\alpha t}, \quad \eta=\gamma \xi_0 e^{\alpha t}, \label{eq:a1c}\\
\xi=-\xi_0 e^{\alpha t}, \quad \eta=-\gamma \xi_0 e^{\alpha t}, \label{eq:a2c}
\end{eqnarray} are two fundamental out going orbits, while two are incoming orbits 
\begin{eqnarray}
 \xi=\xi_0 e^{-\alpha t}, \quad \eta=-\gamma \xi_0 e^{-\alpha t}, \label{eq:a3c}\\
\xi=-\xi_0 e^{-\alpha t}, \quad \eta=\gamma \xi_0 e^{-\alpha t} \label{eq:a4c}
\end{eqnarray}
are images of (\ref{eq:a1c}) and (\ref{eq:a2c}) respectively.

\section{Periodic Orbits in the Neighborhood of Triangular Equilibrium Points}
\label{sec:potp}
The coordinates of the triangular points $L_{4,5}$ are given as \citep{Kushvah2012Ap&SS.337..115K}:
\begin{eqnarray} 
\begin{cases}
x=\frac{q_1^{\frac{2}{3}}}{2}-\mu+(q_1^{\frac{2}{3}} \sigma_1-\sigma_2)\\ 
y=\pm q_1^{\frac{1}{3}}[1-\frac{q_1^{\frac{2}{3}}}{4}+(2-q_1^{\frac{2}{3}}) \sigma_1+\sigma_2]^{\frac{1}{2}} \label{eq:xy}
\end{cases} 
\end{eqnarray}
where $\sigma_1$ and $\sigma_2$ are given as:\begin{eqnarray*}
\sigma_1&=&\frac{1}{3}\left[1-n^2 +\frac{2 \pi c h (b-a)}{a b \{\mu^2 +q_1^{\frac{2}{3}}(1-\mu)\}^{\frac{3}{2}}}\right.\nonumber\\&&\left.+\frac{3 \pi c h ({\lgbya})}{8\{\mu^2 +q_1^{\frac{2}{3}}(1-\mu)\}^2}\right], \label{eq:dl1} \\
\sigma_2&=&\frac{1}{3(1+\frac{5}{2}A_2)}\left[1-n^2+\frac{3A_2}{2}+ \right.\nonumber\\&&\left.\frac{2\pi c h(b-a)}{a b\{\mu^2 +q_1^{\frac{2}{3}}(1-\mu)\}^{\frac{3}{2}}}+\right.\nonumber\\&&\left.\frac{3 \pi c h({\lgbya})}{8\{\mu^2 +q_1^{\frac{2}{3}}(1-\mu)\}^2}\right]. \label{eq:dl2} \end{eqnarray*} 
Now, using parametric values mentioned earlier in addition with triangular point (\ref{eq:xy}) into the equations (\ref{eq:oxx}), (\ref{eq:oyy}) and (\ref{eq:oxy}), we have found the values of $\Omega^0_{xx}$, $\Omega^0_{yy}$ and $\Omega^0_{xy}$. Again, substituting these values in the equation (\ref{eq:ce}), we get another equation  in the variable $\lambda$ and parameter $\mu$ which has a very complicated form. Again, expanding this equation by Taylor's series about the mass parameter $\mu=0$ and taking up to the first order term, we found the following new characteristic equation:
\begin{eqnarray}
 &&\lambda^4+(3.24710+1.84976\mu)\lambda^2+\nonumber\\&&7.38874\mu+2.36659=0.\label{eq:nce}
\end{eqnarray}
Since, the nature of roots of the quadratic equation (\ref{eq:nce}) in $\lambda^2$ depends on the sign of its discriminant $D$ which is given as:
\begin{eqnarray}
 D&=&\left(3.24710+1.84976\mu \right)^2-\nonumber\\&&4\left(7.38874\mu+2.36659\right). \label{eq:dis} 
\end{eqnarray}
Thus, we have the following three cases.

\subsection{When  $D=0; \quad \mu=\mu_c$} 
\label{subsec:dzr}

\begin{figure}
 \plotone{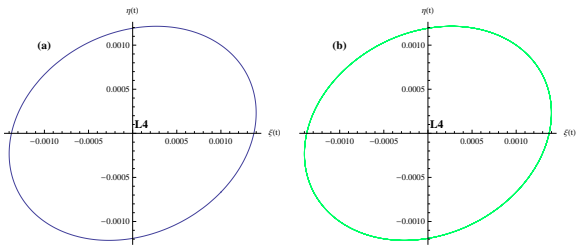}
 \caption{Periodic orbits in the neighborhood of $L_4$: (a)  $0\leq t\leq T(=\frac{2\pi}{1.29655})$  (b)  $0\leq t\leq 5T$ \label{fig3:tp1}}
 \end{figure}

This value of the discriminant directly corresponds to the value of critical mass $\mu_c$ of this problem. In other words
\begin{eqnarray}
 3.42160\left(\mu-5.06476\right)\left(\mu-0.062165\right)=0.\label{eq:crm} 
\end{eqnarray}So, we have the critical mass $\mu_c=0.062165$ because $\mu=5.06476$ is not possible as $0<\mu\leq \frac{1}{2}.$ For this value of $\mu$, the characteristic equation (\ref{eq:nce}) has four imaginary roots $\lambda_{1,2,3,4}$ of equal modulus given as:
\begin{eqnarray}
\alpha i; \quad -\alpha i; \quad \alpha i; \quad -\alpha i,
\end{eqnarray}where $\alpha=1.29655.$ Hence, the solutions of equation (\ref{eq:var}) corresponding to distinct roots will be bounded however due to repeated roots, the general solution of the variational equation (\ref{eq:var}) will be unbounded. The most general solution of variational equation (\ref{eq:var}) can be written as:
 \begin{eqnarray}
 \xi=\left(P_1+P_2 t\right) e^{i\alpha t}+\left(P_3+P_4 t\right) e^{-i\alpha t}, \nonumber\\ 
\eta=\left(Q_1+Q_2 t\right) e^{i\alpha t}+\left(Q_3+Q_4 t\right) e^{-i\alpha t},\label{eq:txie} 
\end{eqnarray}which is of complex type, where $P_j, j=1,2,3,4$ are arbitrary constants whereas  $Q_j, j=1,2,3,4$ are related to them by the equation (\ref{eq:pq1}). That is
\begin{eqnarray}
 Q_j=\left(\frac{\lambda^{2}_j-\Omega^0_{xx}}{2\lambda_j-\Omega^0_{xy}}\right)P_j \label{eq:rel}, j=1,2,3,4.
\end{eqnarray}
The most general real solution of variational equation (\ref{eq:var}) can be written as:  
\begin{eqnarray}
\xi=P \cos{\left(\alpha t+\phi\right)}+P't \sin{\left(\alpha t+\phi'\right)}, \nonumber\\ 
\eta=Q \cos{\left(\alpha t+\psi\right)}+Q't \sin{\left(\alpha t+\psi'\right)},\label{eq:phs}
\end{eqnarray}where  $P', Q', \phi', \psi'$ are related to the four arbitrary integration constants $P, Q, \phi, \psi$ respectively, which are  given as:
\begin{eqnarray}
 &&P=\sqrt{P^2_1+P^2_3}, \  Q=\sqrt{Q^2_1+Q^2_3},\nonumber\\&& \phi=\tan^{-1}{\left(\frac{P_3}{P_1}\right)}, \ \psi=\tan^{-1}{\left(\frac{Q_3}{Q_1}\right)}. 
\end{eqnarray}
Now, to discuss the periodic motion, we have taken $P_2=P_4=0$. Hence, from the equation (\ref{eq:txie}) we have
\begin{eqnarray}
&&\xi=P_1 e^{i\alpha t}+P_3 e^{-i\alpha t},\nonumber\\ 
&&\eta=\gamma_1 P_1 e^{\alpha t}-\gamma_3 P_3 e^{-i\alpha t},\label{eq:xcie}\end{eqnarray}
where $\gamma_1=\frac{Q_1}{P_1}$ and $\gamma_3=\frac{Q_3}{P_3}.$
Again, if $\xi_0$ and $\eta_0$ are the initial displacements then equation (\ref{eq:txie}) at $t=0$ provides:
\begin{equation}
 P_1=\frac{\gamma_3 \xi_0-\eta_0}{\gamma_3-\gamma_1}; \quad P_3=\frac{\gamma_1 \xi_0-\eta_0}{\gamma_1-\gamma_3}.
\end{equation}
The equation (\ref{eq:xcie}) represents the periodic orbits with period $\frac{2\pi}{1.29655}$ of the infinitesimal mass in the vicinity of triangular equilibrium $L_4$ which are depicted in figure (\ref{fig3:tp1}). 
\begin{figure}
 \plotone{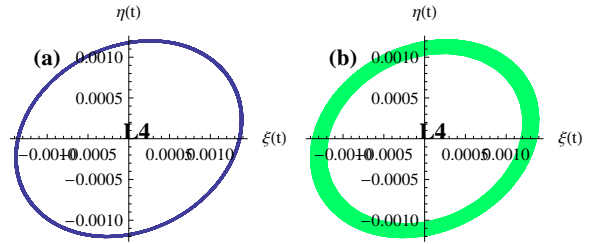}
 \caption{Periodic orbits in the neighborhood of $L_4$: (a)  $0\leq t\leq 200T$  (b) $0\leq t\leq 500T$ \label{fig31:tp1}}
 \end{figure}

Figures (\ref{fig3:tp1}) and (\ref{fig31:tp1}) shows that the orbits of the infinitesimal mass  plotted at different time. From these figures, it can be seen that the orbit have a regular elliptic shape and retaining its path for a long time $t>0$ i.e. if $t=500T$, then the shape of the orbits remains same as for $t=5T$.  It is also  seen that if we draw the parametric equation (\ref{eq:txie}) by assuming $P_1=P_3=0$, then the path of the infinitesimal mass is moving spirally outward with time.

\subsection{When $D>0; \quad  \mu \in(0, \mu_c)$} 
\label{subsec:dps}

In this case, due to positive sign of $D$,  the characteristic equation (\ref{eq:nce}) has four imaginary roots
\begin{eqnarray}
 \lambda_{1,2}=\pm \alpha i; \quad \lambda_{3,4}=\pm \beta i, 
\end{eqnarray}where $\alpha$ and $\beta$ are positive real numbers whose values depend upon the mass parameter $\mu.$ Due to two different imaginary roots $\alpha i$ and $\beta i$, there are two periodic motions of the restricted body with periods $\frac{2\pi}{\alpha}$ and $\frac{2\pi}{\beta}.$ 
 The general complex solution of the variational equation (\ref{eq:var}) in this case is written as:\begin{eqnarray}
 \xi=P_1 e^{i\alpha t}+P_2 e^{-i\alpha t}+P_3 e^{i\beta t}+P_4 e^{-i\beta t}, \nonumber\\ 
\eta=Q_1 e^{i\alpha t}+Q_2 e^{-i\alpha t}+Q_3 e^{i\beta t}+Q_4 e^{-i\beta t},\label{eq:xiet}                                                                                                
 \end{eqnarray}where $P_j, j=1,2,3,4$ are four arbitrary integration constants and $Q_j, j=1,2,3,4$ are related to previous constants by means of relation (\ref{eq:rel}). The more general real solution of (\ref{eq:var}) may be written as follows:

 \begin{eqnarray}
\xi&=&P \cos{\left(\alpha t+\phi\right)}+Q \sin{\left(\beta t+\psi\right)}, \nonumber\\ 
\eta&=&|\gamma_1| P \cos{\left(\alpha t+\phi+\epsilon_1\right)}\nonumber\\&&+|\gamma_3| Q \sin{\left(\alpha t+\psi-\epsilon_3\right)},\label{eq:psh}\end{eqnarray}where $P, Q, \phi, \psi$ are four real arbitrary constants given as:
\begin{eqnarray}
 &&P=\sqrt{P^2_1+P^2_2}, \  Q=\sqrt{P^2_3+P^2_4}, \nonumber\\&& \phi=\tan^{-1}{\left(\frac{P_2}{P_1}\right)}, \ \psi=\tan^{-1}{\left(\frac{P_4}{P_3}\right)}. 
\end{eqnarray} The constants $|\gamma_1|$, $|\gamma_3|$ and  $\epsilon_1$, $\epsilon_3$ are modulus and amplitudes of the complex number $\gamma_j=\frac{Q_j}{P_j}, j=1,3$ respectively.
The equation (\ref{eq:xiet}) represents composed form of two periodic motions known as long periodic motion and short periodic motion with periods $T_{\alpha}=\frac{2\pi}{\alpha}$ and $T_{\beta}=\frac{2\pi}{\beta}$ respectively. The amplitudes of these type of motions are obtained from the integration constants $P_j, j=1,2,3,4$. The looping nature of trajectory of the infinitesimal mass shown in figures is obtained from the two different type of motions that contribute to the perturbed orbit about the equilibrium point. From  figure (\ref{fig6:c2tp3})(a), the time period of the resulting motion is $T=30.4$ (as in our case $\alpha$ and $\beta$ obtained for $0<\mu=0.00001$ are not integers). Again, in case of $\frac {T_{\alpha}}{T_{\beta}}=\frac{s}{k}$, where $s$ and $k$ are integers, the time periods of the resulting orbits is given by the LCM of $T_{\alpha}$ and $T_{\beta}$. For example, time period of the orbits shown in figure (\ref{fig7:int}) are $2\pi$ whereas in frame $(a)$, $\frac{\alpha}{\beta}=\frac{s}{k}$ is equal to $\frac{1}{2}$ and in frame $(b)$ it is equal to $\frac{1}{3}$. Also, the shape of this resulting orbit is bounded and has two loops in former case (figure \ref{fig6:c2tp3}) while in later case (figure \ref{fig7:int}$(a \& b)$) have one and two loops respectively. The loops in a quasi-periodic orbit, are very helpful to inspect the order of resonance.

Now, to study the separate motion of the infinitesimal body, we have considered the motion (\ref{eq:psh}) with $Q=0.$ That is
\begin{eqnarray}
 &&\xi=P \cos{(\alpha t+\phi)}, \nonumber\\ 
&&\eta=|\gamma_1| P \cos{(\alpha t+\phi+\epsilon_1)},\label{eq:spsh}
\end{eqnarray}
which represents a one parameter family of ellipse. Now, eliminating parameter $t$ by trigonometric simplification in between both the equation of (\ref{eq:spsh}), we get
\begin{eqnarray}
 \xi^2+\frac{\eta^2}{|\gamma_1|^2}-\frac{2\xi \eta}{|\gamma_1|}{\cos\epsilon_1}=P^2\sin^2{\epsilon_1}.\label{eq:telp}
\end{eqnarray}
The equation (\ref{eq:telp}) represents an ellipse which is bounded by the region $\xi=\pm P$ and $\eta=\pm |\gamma_1|P.$
  \begin{figure}
   \plotone{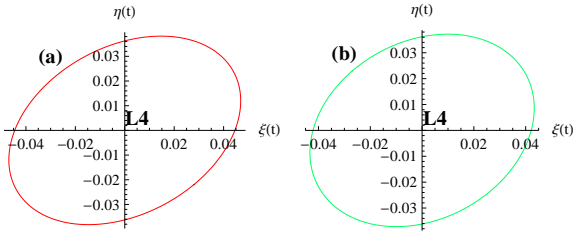}
  \caption{Periodic orbits in the neighborhood of $L_4$: (a)  $0\leq t\leq T_{\alpha}(=\frac{2\pi}{1.05102}$  (b)  $0\leq t\leq T_{\beta}(=\frac{2\pi}{1.46372}$\label{fig4:c2tp1}}
  \end{figure}
 \begin{figure}
  \plotone{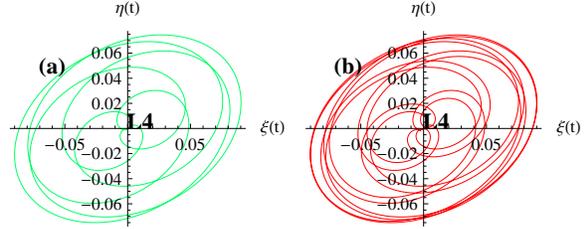}
 \caption{Orbits in the neighborhood of $L_4$ when  $\frac{\alpha}{\beta}=\frac{s}{k}$ and $s, k$ are not integers: (a)  $0\leq t\leq T(=30.4)=7.1T_{\beta}$  (b)  $0\leq t\leq 2T$ \label{fig6:c2tp3}}
   \end{figure}
 \begin{figure}
   \plotone{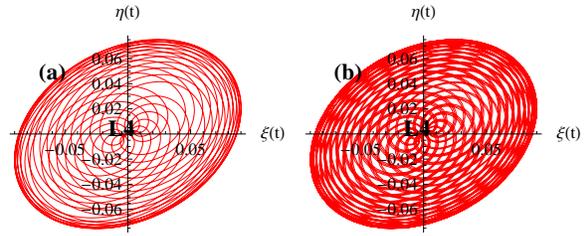}
  \caption{Orbits in the neighborhood of $L_4$  when  $\frac{\alpha}{\beta}=\frac{s}{k}$ and $s, k$ are not integers: (a) $0\leq t\leq 5T$  (b)  $0\leq t\leq 30T$ \label{fig7:c2tp4}}
   \end{figure}
\begin{figure}
   \plotone{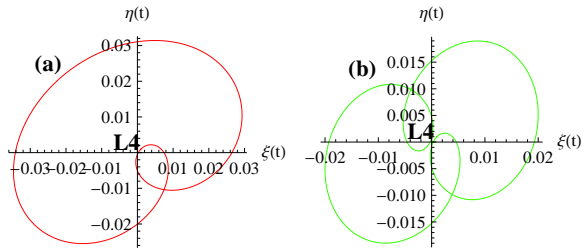}
  \caption{Orbits in the neighborhood of $L_4$  when  $\frac{\alpha}{\beta}=\frac{s}{k}$ and $s, k$ are integers: (a)  $0\leq t\leq 2\pi$  (b)  $0\leq t\leq 2\pi$ \label{fig7:int}}
   \end{figure}
In Figure (\ref{fig4:c2tp1}) which is plotted at $\mu_c=0.00001$, frames $(a)$ and $(b)$ show that the orbits of infinitesimal mass in the vicinity of $L_4$ with periods $\frac{2\pi}{1.05102}$ and $\frac{2\pi}{1.46372}$ correspond to roots $\lambda_{1,2}=\pm 1.05102 i$ and $\lambda_{3,4}=\pm 1.46372 i$ respectively. From this figure, it is also clear that orbits are elliptic and bounded for a long time $t>0$(as in figure \ref{fig31:tp1}) . On the other hand, we have depicted the shape of composed form of orbits of two periodic motions in figures (\ref{fig6:c2tp3}) and (\ref{fig7:c2tp4}) at different time $t$ when $\frac{\alpha}{\beta}\neq\frac{s}{k}$. Figure (\ref{fig7:int}) for the case when $\frac{\alpha}{\beta}=\frac{s}{k}$, where $s$ and $k$ are integers i.e. in frame (a) $\frac{s}{k}=\frac{1}{2}$ and in frame $(b)$  it is $\frac{1}{3}$. From   figure (\ref{fig6:c2tp3})$(a)$, it is clearly seen that there is a second order resonance in the quasi-periodic orbit (\ref{eq:xiet}). 

\subsection{When $D<0; \quad  \mu \in(\mu_c, \frac{1}{2})$} 
\label{subsec:dng}
In this case the four roots of the characteristic equation (\ref{eq:nce}) have the form
 \begin{eqnarray}
 &&\lambda_{1}= \alpha+i\beta; \ \lambda_{2}= \alpha-i\beta; \nonumber\\&& \lambda_{3}= -\alpha+i\beta; \ \lambda_{4}= -\alpha-i\beta, 
\end{eqnarray}where $\alpha$ and $\beta$ are positive real numbers depending on the mass parameter $\mu.$ These four roots are responsible for the motions of infinitesimal body in the basin of triangular equilibrium points. The motion is periodic with period $\frac{2\pi}{\beta}$ however the amplitude is an exponential function of parameter $t$ and $\alpha.$ In this case the most general complex solution of the equation (\ref{eq:var}) may be written as:
\begin{eqnarray}
 \xi=P_1 e^{\lambda_1 t}+P_2 e^{\lambda_2 t}+P_3 e^{\lambda_3 t}+P_4 e^{\lambda_4 t}, \nonumber\\ 
\eta=Q_1 e^{\lambda_1 t}+Q_2 e^{\lambda_2 t}+Q_3 e^{\lambda_3 t}+Q_4 e^{\lambda_4 t},\label{eq:xie3}                                                                                                
 \end{eqnarray}where constants $Q_j, j=1,2,3,4$ related to four arbitrary integration constants $P_j, j=1,2,3,4$ by mean of (\ref{eq:rel}). Also, the most general real solution of (\ref{eq:var}) have the form:
\begin{eqnarray}
 \xi&=&e^{\alpha t}\left(P_1 \cos{\beta t}+P_2 \sin{\beta t}\right)+\nonumber\\&&e^{-\alpha t}\left(P_3 \cos{\beta t}+P_4 \sin{\beta t}\right), \nonumber\\ 
\eta&=&e^{\alpha t}\left(Q_1 \cos{\beta t}+Q_2 \sin{\beta t}\right)+\nonumber\\&&e^{-\alpha t}\left(Q_3 \cos{\beta t}+Q_4 \sin{\beta t}\right),\label{eq:rxi3}
 \end{eqnarray}
where real constants $Q_j, j=1,2,3,4$ are the function of arbitrary real constants $P_j, j=1,2,3,4.$ Now, if we take $P_3=P_4=0$ in equation (\ref{eq:rxi3}) then we found that orbit is of asymptotic type which moves away from the triangular equilibrium point with time $t$. The period and amplitude of this orbit are $\frac{2\pi}{\beta}$ and $e^{\alpha t}$ respectively. On the other hand, if we assume $P_1=P_2=0$ in equation  (\ref{eq:rxi3}) then the motion of infinitesimal body is similar to previous one provided amplitude $e^{-\alpha t}$ should decrease with the time.
 \begin{figure}
 \plotone{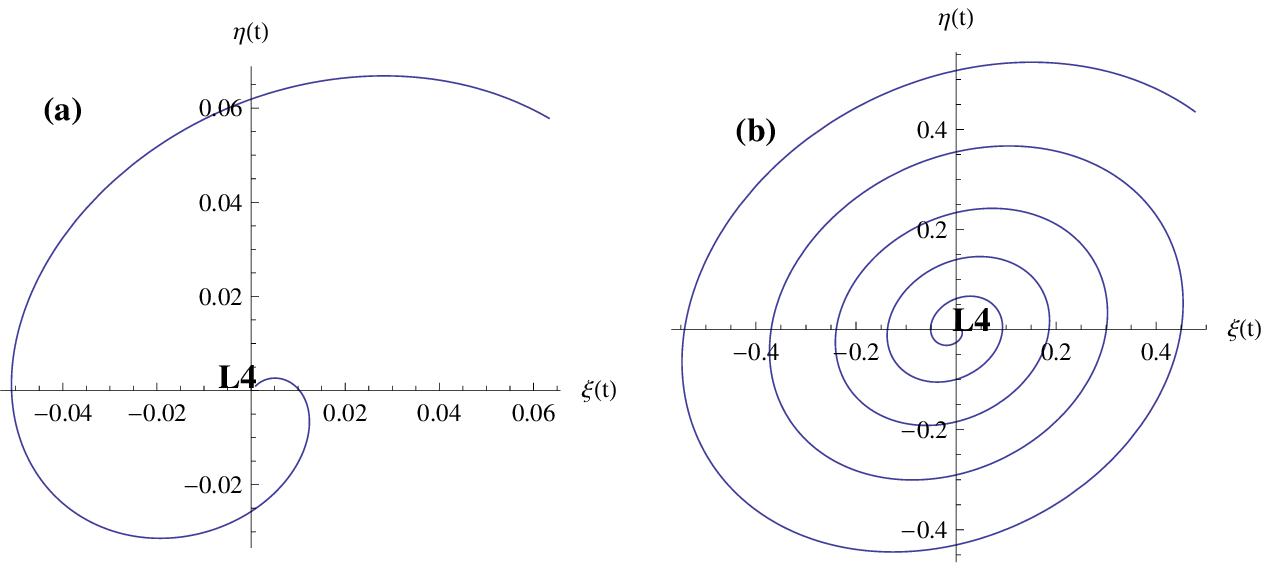}
 \caption{Orbits in the neighborhood of $L_4$: (a)  $0\leq t \leq T(=\frac{2\pi}{1.301243})$  (b) $0\leq t \leq5T$ \label{fig8:c3tp1}}
 \end{figure}
\begin{figure}
\plotone{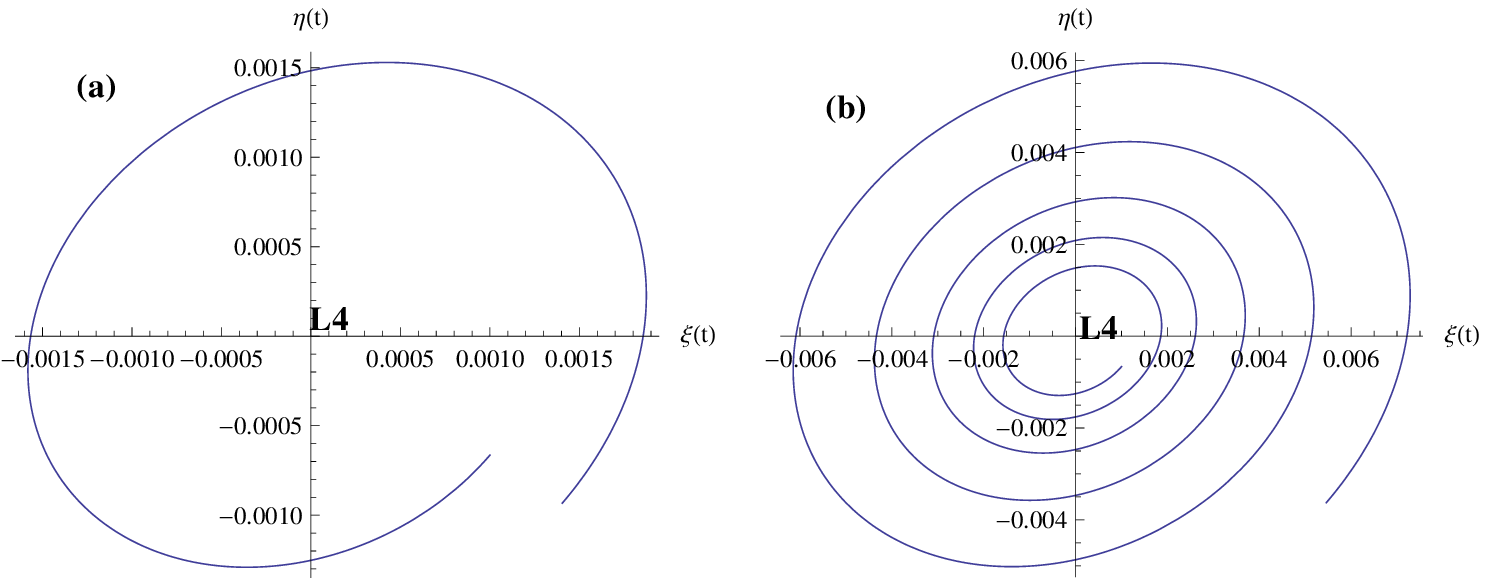}
 \caption{Orbits in the neighborhood of $L_4$: (a)  $0\leq t\leq T(=\frac{2\pi}{1.301243})$  (b) $0\leq t\leq5T$ \label{fig9:c3tp2}}
 \end{figure}
 
In figure (\ref{fig8:c3tp1}) which is plotted at $\mu_c=0.0651$, frame $(a)$ and $(b)$ show that the path of infinitesimal mass is of asymptotic type. It is also seen that the orbit is going away from center $L_4$ with time $t$. Figure (\ref{fig9:c3tp2}) drawn by taking $P_3=P_4=0$ shows that the amplitude of the orbit increases with time.

\section{Stability Criteria}
\label{sec:stb}
 Since, It is already known that to find the orbital parameters of the infinitesimal mass, we require position $(x,y)$ as well as velocity co-ordinates $(\dot x,\dot y)$ at different instant i.e. we have a four dimensional phase space $(x,y,\dot x,\dot y)$. Again, if we take $x, y$ and $\dot x$ as our three variables then the other one $\dot y$, is obtained with the help of Jacobi integral equation (\ref{eq:ji}) provided that the value of Jacobi constant at time $t=0$ is known. Now,  we define plane $y=0$ in the remaining three dimensional space $(x, y, \dot x)$, then the plot $x$ Vs $\dot x$ at every time when the trajectory intersects the plane $y=0$ in a particular direction $\dot y>0$ gives the PSS \citep{Murray2000ssd..book.....M}. The PSS is a very helpful to determine the region of stability of the trajectory. The smooth well defined islands represent the trajectory is likely to be regular whereas the islands liberate an exact resonance in between the mean motions of infinitesimal mass and perturber. In other words, appearance of such islands is a characteristic of resonant motion. In that cases a mean motion resonance of the form $a_1+a_2:a_1$, where $a_1$ and $a_2$ are integers, provides $a_2$ islands.  Any fuzzy distribution of the points in the PSS indicates that the trajectory is chaotic. According to KAM theory, when a curve shrink down to a point then this point corresponds to  a periodic orbit. In general, the regular regions in the PSS represents the stability regions but inside the regular region there is a negligible chaotic region whereas out side of that region motion will always be unstable.
\begin{figure} 
\plotone{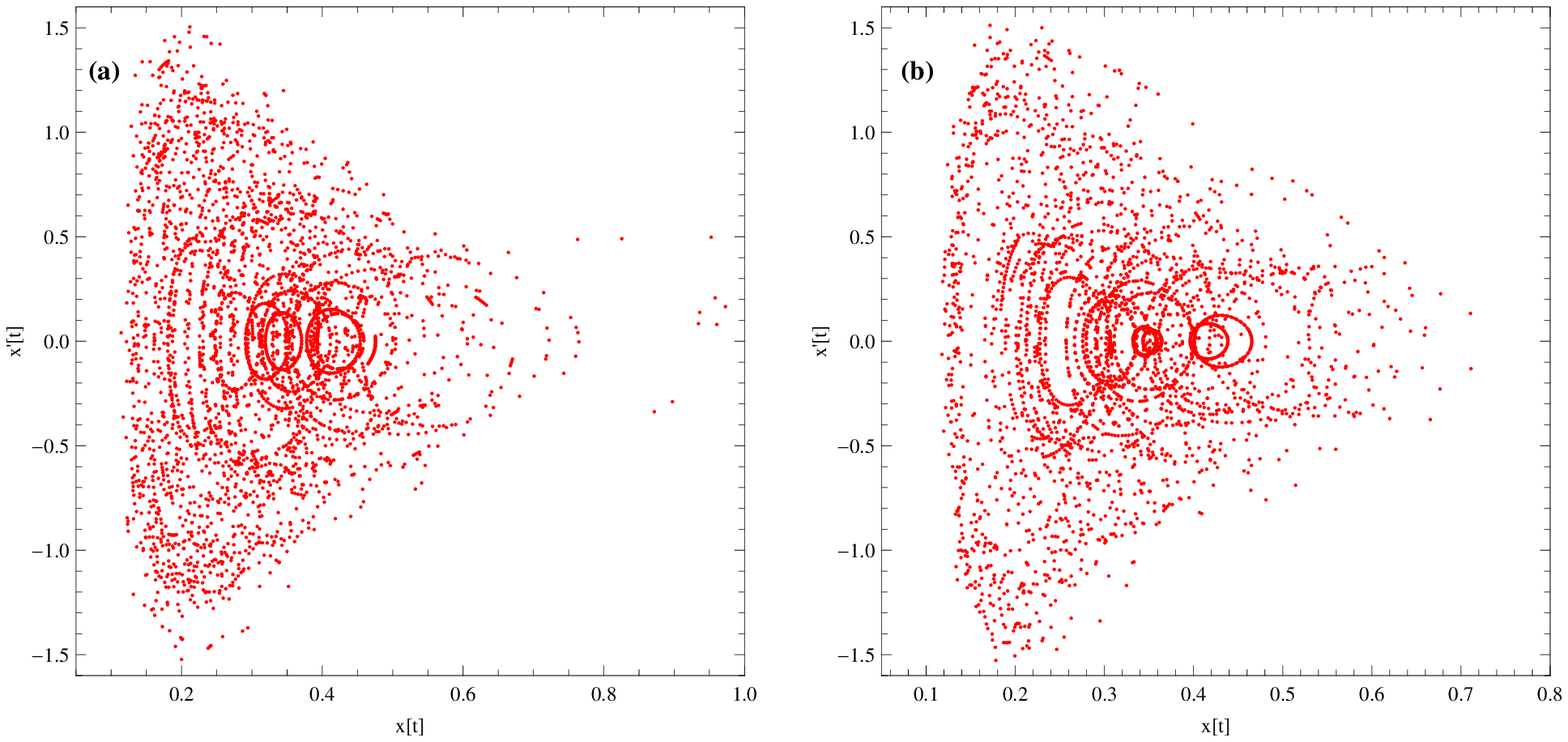}\\\plotone{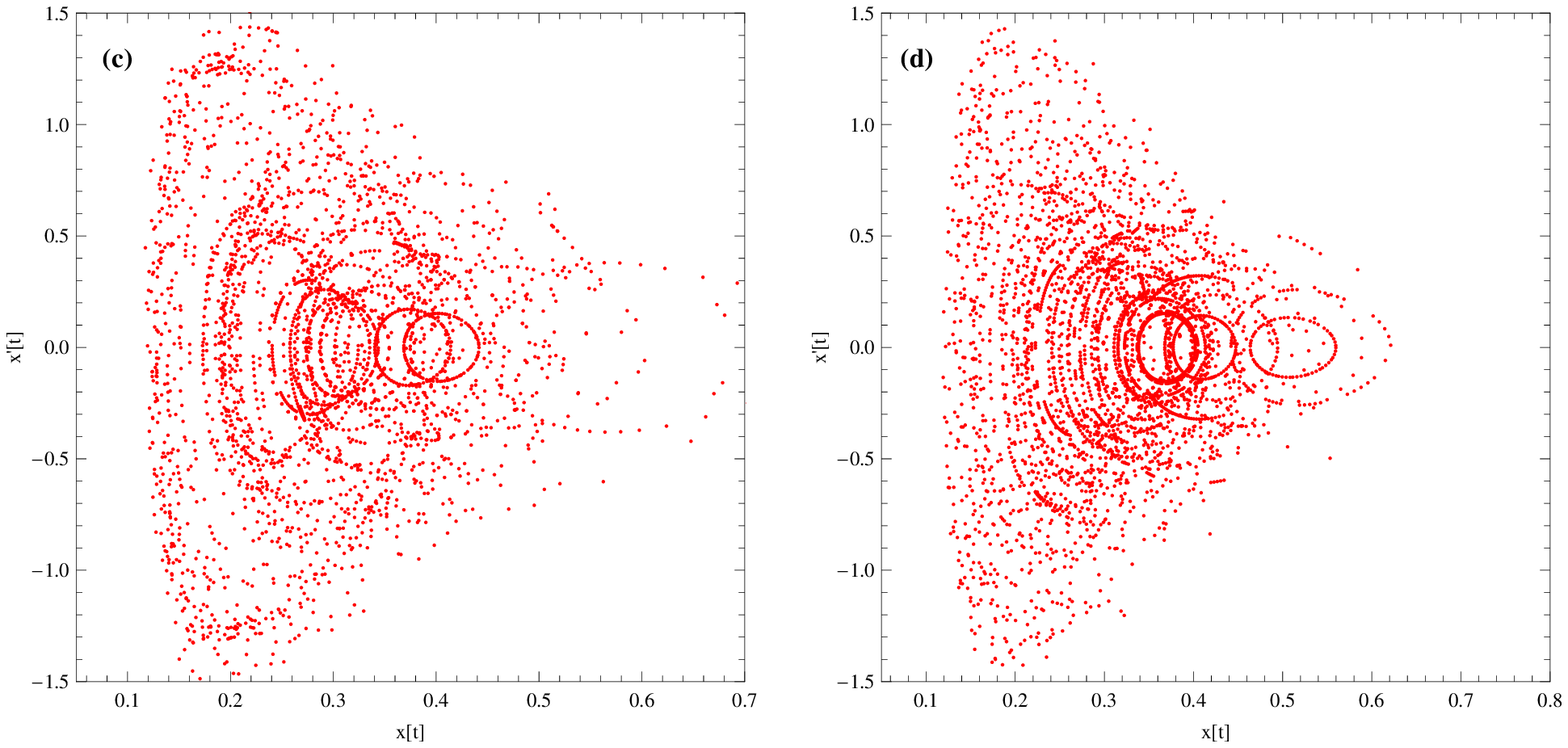}\\\plotone{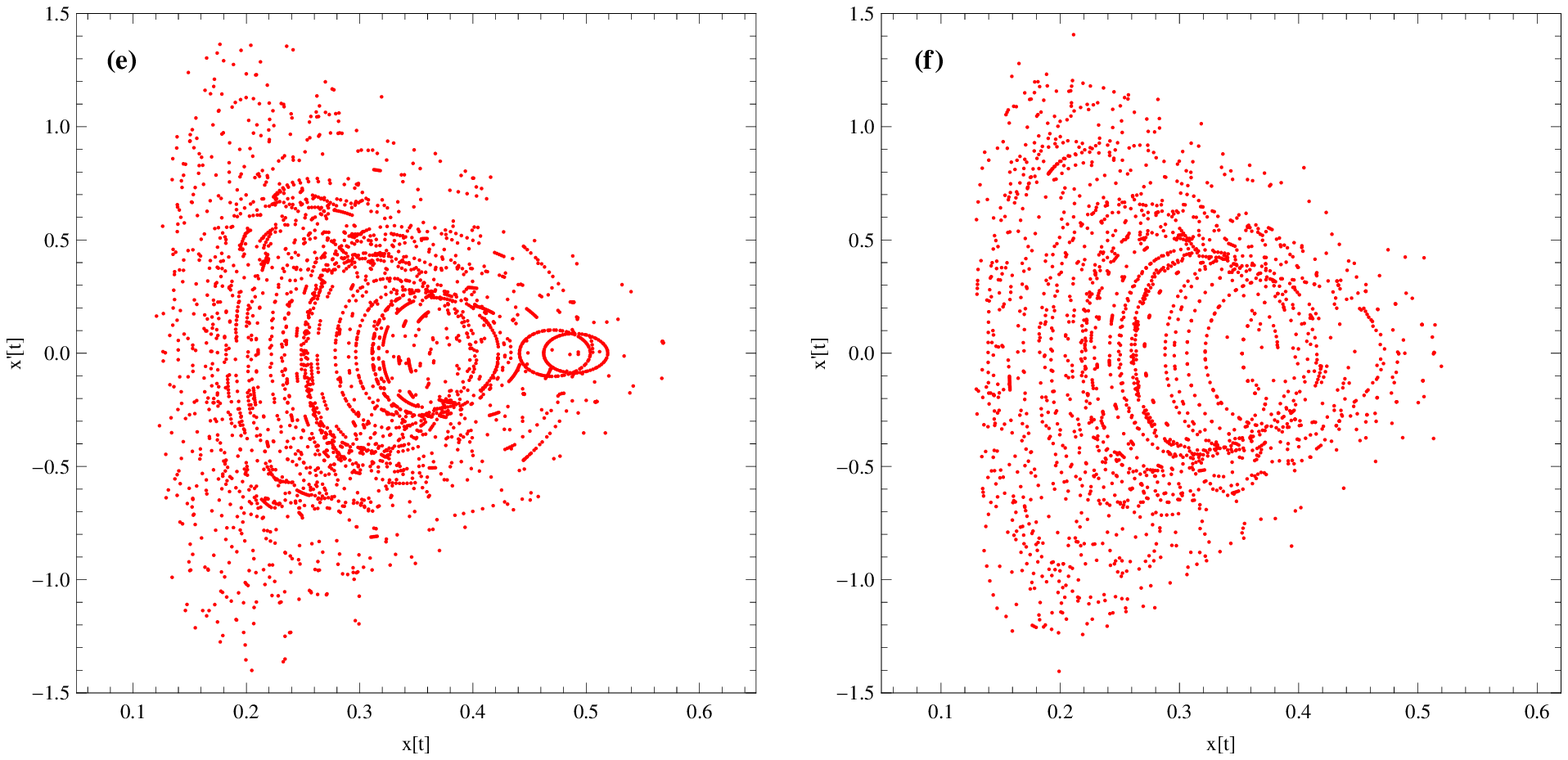}\\\plotone{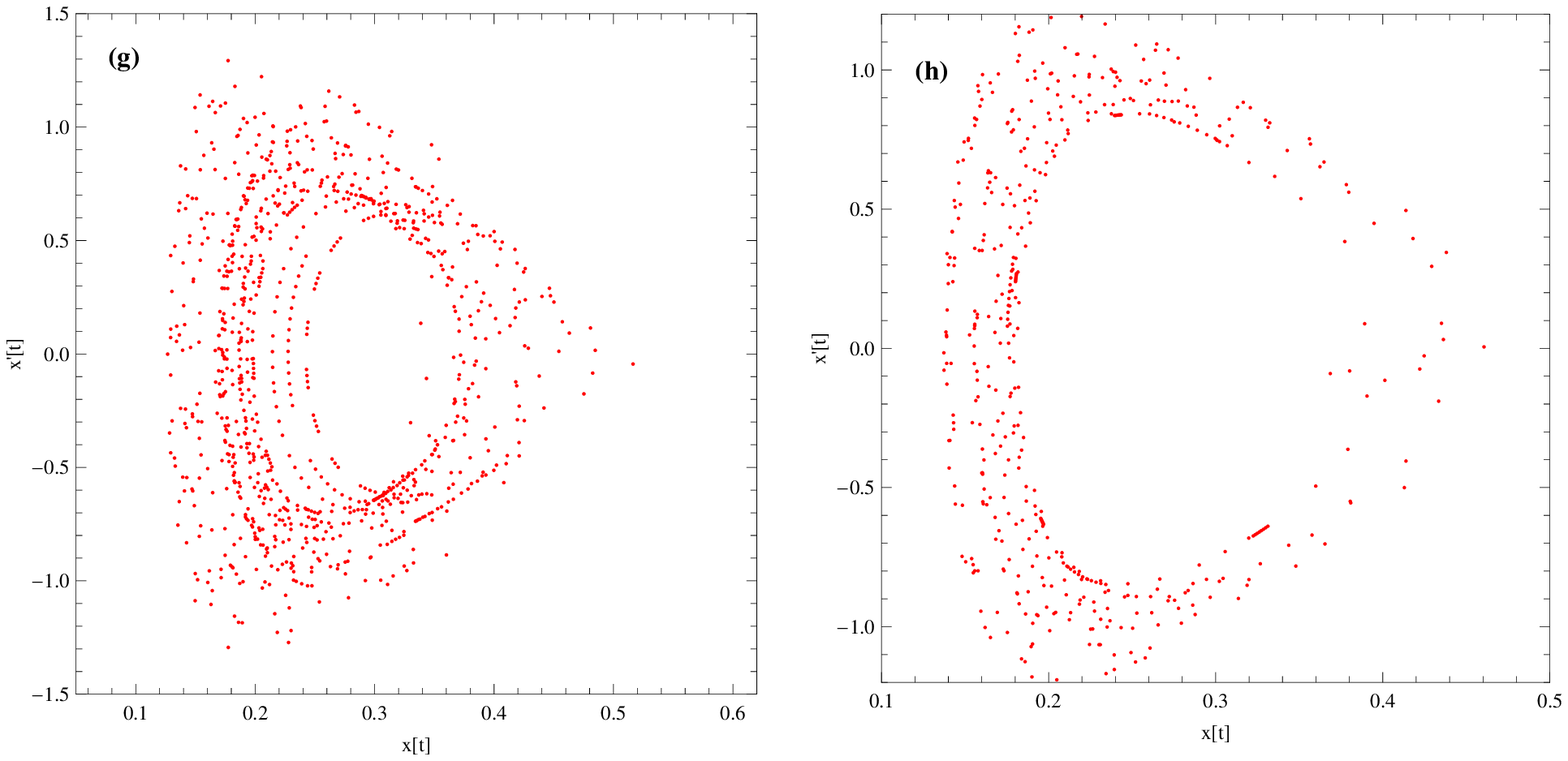}
\caption{Poincar\'{e} surfaces of section at different values of Jacobi constant: $(a) C=2.70, (b) C=2.80, (c) C=2.90, (d) C=3.00, (e) C=3.10, (f) C=3.20, (g) C=3.30$ and $(h) C=3.35$ \label{fig:ps}}
 \end{figure}
\begin{figure} 
 \plottwo{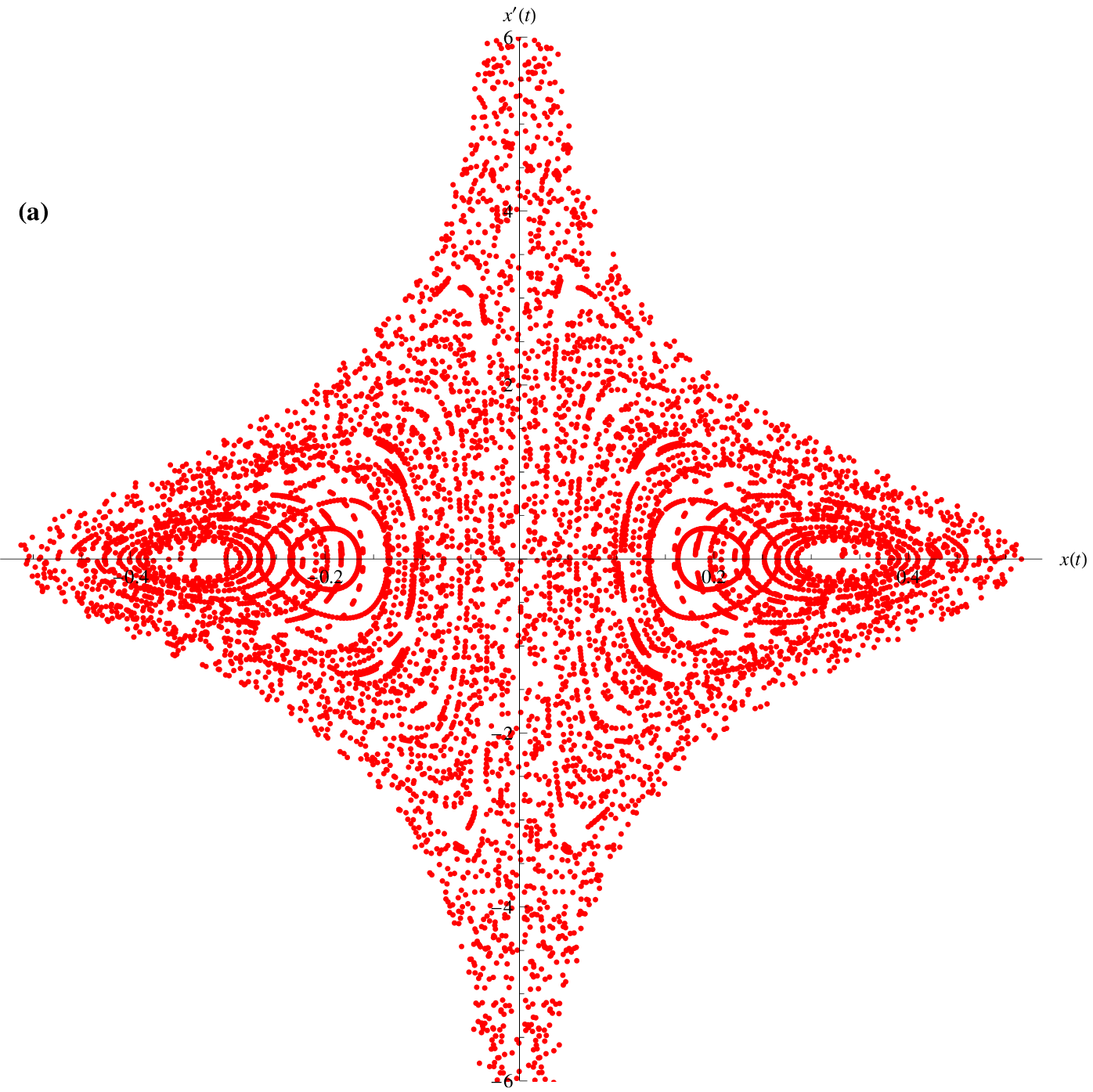}{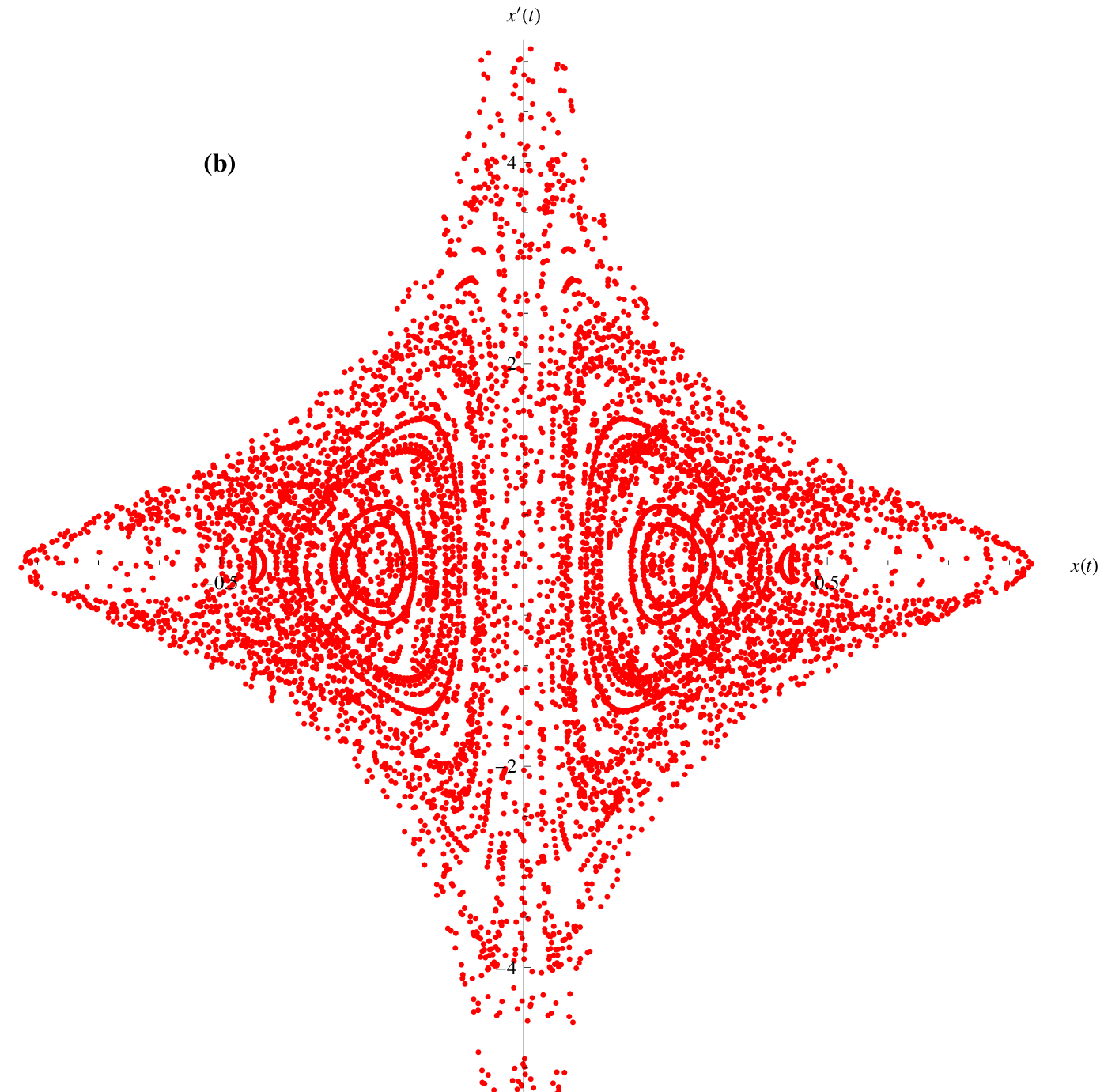}
\caption{Poincar\'{e} surfaces of section at $C=3.10$: (a) when $q_1=0.75$ and $A_2=0.0$ (b) when $q_1=1.0$ and $A_2=0.0025$ \label{fig10:ps2}}
 \end{figure}

Figure (\ref{fig:ps}) shows the PSS at various values of Jacobi constant $C$. From frames (a) to (d), it is clear that the regular islands correspond to the stability region get increases from $C=2.70$ to $C=3.00$ while in frames (e) to (h), it seems to be disappear gradually from $C=3.10$ to $C=3.35$. Also, the fuzzy distribution of the points in the PSS  decrease with an increase in the value of $C$. In other words, extent of chaos decreases with $C$. The existence of chaotic regions does not mean that the trajectory is unbounded but they cover a large area of the plane $(x,\dot x)$. For example, frames (g \& h) shows that the orbits can remain confined for a long time at a particular value of $C$. The PSS in figure (\ref{fig10:ps2}), plotted for the initial conditions $x$ and $\dot x$ lie in the interval $(-0.4,0.4)$ in which frames (a) and (b) correspond to two sets of parameters. Clearly, It can be seen that there are two islands in the plane $(x,\dot x)$ corresponds to each of the initial conditions within the range which indicates that orbits have mean motion resonance of second order.
\section{Conclusion}
\label{sec:conc}
We have studied different aspects of the periodic motions of infinitesimal mass in the neighborhood of equilibrium points for the mass parameter $0<\mu \leq \frac{1}{2}$ and found the periodic orbits in addition to some other kind of orbit like hyperbolic, asymptotic etc. We have seen that some of periodic orbits are bounded and have elliptic shapes while some are unbounded and have spiral shapes. We have also found that the motion of infinitesimal mass is affected by radiation factor as well as oblateness coefficients of the primaries respectively. The stability of the motion is discussed with the help of PSS method and found that the stability regions first increases with Jacobi constant from $C=2.70$ to $C=3.00$ and then it decreases gradually from $C=3.10$ to $C=3.35$. Thus, we conclude that the perturbation factors and the presence of disk  play a very significant role for the periodic motion of the bodies in space. 
 
 \acknowledgements{We are thankful to the Department of Science and Technology, Govt. of India for providing financial support through the SERC-Fast Track Scheme for Young Scientist [SR/FTP/PS-121/2009]. We are also thankful to IUCAA Pune for partial support to visit library and to use computing facility.}
\bibliographystyle{spbasic} 

\end{document}